\documentclass[]{article}

\def\ben{\begin{equation}}
\def\een{\end{equation}}
\def\bea{\begin{eqnarray}}
\def\eea{\end{eqnarray}}
\input amssym.def
\input amssym.tex
\begin{document}

\hfuzz=100pt
\title{Some comments on Gravitational Entropy and the Inverse Mean Curvature Flow}
\author{G. W. Gibbons
\\
D.A.M.T.P.,
\\ Cambridge University, 
\\ Silver Street,
\\ Cambridge CB3 9EW,
 \\ U.K.}
\maketitle

\begin{abstract} 
The  Geroch-Wald-Jang-Huisken-Ilmanen
approach to the positive  energy problem to
may be extended to give
a negative lower bound for the mass
of  asymptotically Anti-de-Sitter spacetimes 
containing horizons with exotic topologies having
ends or infinities  of the form $\Sigma_g \times {\Bbb R}$,
in terms of the cosmological constant. We also show how the 
method gives a lower bound for for the mass of 
time-symmetric initial data
sets for black holes with vectors and scalars in terms
of the mass, $|Z(Q,P)|$
of the double extreme black hole with the same charges.
I also give a lower bound for the area of an apparent horizon,
and hence a lower bound for the entropy in terms of the same function
$|Z(Q,P)|$.  
This shows that the  so-called
attractor behaviour extends beyond the static
spherically symmetric case.
and underscores the general importance of the function $|Z(Q,P)|$.
There are hints that higher dimensional generalizations may involve the Yamabe
conjectures.
\end{abstract}

\section{Introduction}

Recently Huisken and Ilmanen \cite{HI1} \cite{HI2}
have made mathematically rigourous
an old idea of Geroch's \cite{G}
for proving the positive mass
theorem using the inverse mean curvature flow. It was realized by 
Jang and Wald \cite{JW}
soon after Geroch's suggestion that if the method could be made to work
it would yield a lower bound for the mass of an 
asymptotically flat spacetime in terms of the area of any apparent horizon.
The inverse mean curvature flow is not in general
smooth and has jumps. Nevertheless Huisken and Ilmanen are able to
show that despite the jumps the basic idea of Geroch goes through
because the functional that he introduced is monotonic through a jump.

In this note I want to point out that Geroch's formal
 argument may be 
extended to cover exotic black holes with non-trivial
topology which occur in theories with a negative cosmogical term
and to cover black holes with scalar and abelian
vector fields. That is locally the flow will be monotonic. Ir seems reasonable to believe that the rigorous methods of Huisken and Ilmanen
can then be extended to this setting to give a genuine proof.
Some of the results below have been known to me
for many years but in the absence of a proper proof
for the existence of the Geroch flow I hesitated 
to publish them. In addition they were of limited
cosmological or astrophysical interest.
Publication seems more appropriate now that a proper proof is at
hand and in the light  of the prominent role that
 these black holes play 
in current attempts to derive black hole entropy 
in terms of microstates and in applications to 
conformally invariant gauge theories via the AdS/CFT correspondence.
It is precisely rigorous bounds on the classical entropy
of initial data sets which could substantiate some of the ideas 
currently labelled 
with the name "holography" 

Asymptotically flat black holes
with scalar $\phi$ and vector fields carrying electric charges $Q$
and magnetic charges $P$
arise in supergravity and supergravity theories. Their properties are governed by a function $V(\phi, Q,P)$ quadratic in charges and depending on the 
 manifold in which the scalars take their values. 
We  define a function $Z(Q,P)$ which is
the value of $\sqrt {V(\phi,Q,P)}$ at its critical point. 
The bounds we obtain on the entropy
and mass may be expressed entirely in terms of $Z(Q,P)$.
In supersymmetric theories $Z(Q,P)$ is related to the central
charges of the theory but it may be defined more generally.
Previous work has largely been in the context of static
spherically symmetric 
solutions ( see \cite{GKK,FGK}and references therein) . The main point 
being made here is that $V(\phi, Q, P)$
and $Z(Q,P)$ retain their importance when considering time dependent and non-spherically symmetric situations. This is completely
consistent with their microscopic interpretation in terms
of states of D-branes. Moreover it provides a remarkable link 
between  thermodynamic ideas and global diferential geometry
which may well extend much further. Some links bewteen
 entropy, complexity, and gravitional action 
in the context of hyperbolic geometry have already been made \cite{GG1,GG2}.

These general thermodynamic ideas extend to all dimensions.
However at present the inverse mean curvature flow techniques are restricted
to 3+1 dimensional physics. In the last section of this paper I indicate
the difficulties one encounters. It appears that overcoming them may 
involve the Yamabe conjectures, a subject which has also been applied to 
gravitational entropy in a cosmological context \cite{GG3,GG4}.
In fact the calculations in \cite{GG3,GG4} involved the
entropy of the matter in a self consistent
solution of the Einsten equations resulting in an  Einstein static universe,
$ESU_4 \equiv {\Bbb R} \times S^3$. Since $ESU$ is both
the universal cover
of the conformal compactification of four-dimensional
Minkowski spacetime ${\Bbb E} ^{3,1}$ and the conformal boundary
of five-dimensional  Anti-De-Sitter spacetime $AdS_5$, it is not inconceivable
that the AdS/CFT corepondence may entail the Yamabe conjectures
in some fundamental way.

\section{Time Symmetric Initial data sets}
We shall consider three-dimensional initial data sets $\{\Sigma, g_{ij}\}$
for general relativity with a cosmological term.
The arguments will also go through
if we merely demand that the initial data set is maximal
(i.e. if the  trace $
K=g^{ij}K_{ij}$  of the second fundamental form $K_{ij}$
vanishes) but since this is an entirely straightforward generalization
which introduces no new features, I   shall not say more about it here.

The Ricci scalar $R$  of the initial metric is constrained
to satisfy
\ben
R= 2\Lambda +16\pi T_{00} 
\een
where $T_{00}$ is the energy density of the matter
and $\Lambda$ is the cosmological constant.

A basic model example
is given by
\ben
ds^2={ dr^2 \Delta} +r^2 d \omega^2_k 
\een
where $ d \omega^2_k $ is the metric of constant Guassian
curvature $=k=0, \pm 1$ and 
\ben
\Delta= k-{ 2M \over r} -{\Lambda \over 3} r^2.
\een
This three-metric is just 
that induced on the
constant time  hypersurfaces of the Kottler
or 
Schwarzschild-De-Sitter static
spacetime
 \ben
ds^2 = - \Delta dt^2 + g_{ij} dx^i dx^j.
 \een
We are mainly interested in the cases when the
two-surfaces of constant $r$ are closed and orientable
with genus $g$. Thus if $k=1$ we have two-spheres
with $g=0$, if $k=0$ we have tori with $g=1$ and if $k=-1$ we have
$g\ge2$. We may view this last case as $H^2/\Gamma$ where 
$H^2$ is two-dimensional hyperbolic space and $\Gamma$ is a suitable
discrete subgroup of its isometry group $S(2,1)$.

The interesting new cases occur when $
\Lambda <0$ \cite{M} . We then let $\Lambda = - {3\over a^2}$
and thus
\ben
\Delta= { r^2 \over a^2} + k-{ 2M \over r}. 
\een

If $M=0$ we have Anti-de-Sitter spacetime, $AdS_4$
, or a quotient of it by a discrete group $\Gamma$.
We may think of $AdS_4$ as a quadric in ${\Bbb E}^{3,2}$ :
\ben
(X^0)^2 + (X^4)^2 -(X^1)^2 -(X^2)^2-(X^3)^2= -{ 3 \over \Lambda}.
\een
The isometry group of $AdS_4$ is $SO(3,2)$. 
The three (locally) static forms of the metric correspond to
the three different types of one-parameter subgroups of $SO(2,1) \subset
SO(3,2)$ acting say on the coordinates $(X^0,X^4, X^3)$.

Thus the globally static case $k=1$ corresponds to rotations
in the $X^0, X^4$ 2-plane. The case $k=-1$ corresponds
to boosts in the $X^0, X^3$, and the discrete group
$\Gamma \subset SO(2,1)$ acting 
in the $X^4,X^1,X^2$ 3-plane. We have a killimg horizon
and with respect to this Killing field
$AdS_4$ has temperature ${ 1\over 2\pi} \sqrt {-\Lambda \over 3}$.

The case $k=0$ corresponds to
null rotations and there is a degenerate Killing horizon
If one identifies the horizon 
to get a torus
the identfications are also null rotations. 
However these identifications do not act freely on 
the horizon and introduce orbifold singularities there.

Now if we pass to the the Kottler solution
 when $M\ne0$
we find that if $k\ge0$ then $M>0 $ guarantees the 
existence of a regular apparent horizon
of area $A= 4 \pi r^2_H$ where $r_H$ is the outermost root
of $\Delta(r_H)=0$. In fact 
\ben
2M={r^3 _H \over a^2} +  k r_H. 
\een
If $k=0$  and $M=0$ then the spatial section
$\Sigma$ is complete and  has a cusp
at at $r=0$ which is  an  infinite spatial distance.
The static Kottler spacetime has a degenerate Killing horizon at the cusp
in this case.
If $k=-1$ we have
\ben
2M= {r_H^3 \over a^2} +  k r_H. 
\een
Thus $M$ can be negative but not too negative. If
\ben
M> { 1 \over \sqrt{27 a^2} }
\een
then we have a regular apparent horizon
of area $ A=4\pi|1-g| r^2_H$. The case of equality
\ben
M= { 1 \over \sqrt{27 a^2}  }
\een
gives a cusp at
\ben
r={ 1\over \sqrt{3 a^2}}.
\een

These observations suggest a general 
lower bound, valid for all values of $k$
 of the form

\ben
2M \ge {r_H^3 \over a^2} +  k r_H. 
\een
 
We shall shortly argue that precisly this lower bound is
obtainable form the inverse mean curvature flow.
In particular this seems to imply that the 
the negative mass topolologicaly non-trivial
black hole are classically stable. Moreover the degenerate
zero tempertaure limiting solution should be qunatum mechanically
stable even though it is not a BPS state, that is,  
it has no Killing spinors and hence admits no supersymmetry.

\section{The Geroch flow}

The argument initiated by Geroch \cite{G}, 
extended by Wald and Jang \cite{JW} and completed
by Ilmanen and Huisken \cite{HI1,HI2} goes roughly as follows.
One considers a family of level sets
$S_s$ which are smoothly  immmersed two-surfaces
each with metric $h_{ab}$, second fundamental form $p_{ab}$
of area $A(s)$ mean curvature $p=h^{ab} p_{ab}$ and Gaussian curvature $K_G$
evolving in $s$ according to the inverse mean curvature flow equation so the 
velocity $v_n$ of the surface $S_s$ along its normal ${\bf n}$
is given by 
\ben
v_n= { 1\over p}. 
\een
It follows that
\ben
{ dA \over ds} = A.
\een
One associates with each surface the function:
\ben
f(s)= \int_{S_s}(4K_G-p^2 - { 4 \Lambda \over 3} )   dA
\een
and finds that
\ben
{ d (f A^{ 1\over 2} ) \over ds}
 \ge  A^{ 1\over 2} \int_{S_s} 16 \pi T_{00}.
\een

Equality is possible if and only if
the trace free part of the extrinsic curvature of the surfaces $S(s)$
vanishes and $p$ is constant on the surfaces.
Now start the flow from the outermost  apparent horizon (which of course
may not be connected) on which $p=0$. One assumes that
the flow reaches infinity 
near which the metric behaves like the model example.
If the surfaces tending to $r={\rm constant }$ surfaces,
one finds that  
\ben
\lim f A^{ 1\over 2}  = 64 \pi ^ { 3\over 2} M.
\een
If the matter energy density $T_{00}$ is non-negative
one may integrate the inequality to get
\ben
 64 \pi ^ { 3\over 2} M \ge fA^{ 1\over 2} _H.
\een

One now evalutes $ f A^{ 1\over 2} $ on the apparent horizon
which in this case will have vanishing mean curvature, $p=0$
to obtain the desired inequality. 

Actually the  term in $f$ involving the cosmological constant
was first introduced in \cite{BGH} and the results of \cite{HI2} may need extending to cover this case. I shall assume in what follows that that is possible.

\section{Charged Black Holes}

If the energy density takes a particular form
then stronger inequalities may be obtained.
Thus in Einstein-Maxwell theory the model
metric is the Reissner-Nordstrom-De-Sitter
for which
\ben
\Delta= k-{ 2M \over r} -{\Lambda \over 3} r^2 + { Z^2 \over r^2}.
\een
where
\ben
Z^2= Q^2 + P^2,
\een
and $P$ and $Q$ are the electric and magnetic charges.
One expects
\ben
2M \ge {r_H^3 \over a^2} + k r_H + { Z^2 \over r_H^2} . 
\een
To obtain this one follows Jang \cite{J} and notes that
\ben
T_{00}= { 1\over 8 \pi} ( {\bf E}^2 + {\bf B}^2 ),
\een
where $\bf E$ and $\bf B$ are the electric and magnetic fields.
They are divegence free with respect to the metric $g_{ij}$
on $\Sigma$:
\ben
{\nabla}. {\bf E}=0={\nabla}.{\bf B}.
\een
The electric charge $Q$ inside a level set is
is given by
\ben
Q= { 1 \over 4\pi } \int_{S_s} E_n dA
\een
where $E_n$ is the normal component
of the electric field $\bf E$
and there is a similar expression for the magnetic charge $P$.

Now
\ben
T_{00}\ge  { 1\over 8 \pi} ( { E_n}^2 + {B_n}^2 ),
\een
and use of the Schwartz Inequality gives:
\ben
 \int_{S_s} 16 \pi T_{00} \ge 32 \pi^2{ Z^2 \over A(s)}.
\een
One now integrates the inequality to get the desired result.

\section{Entropy and Attractors}

In this section we shall extend the argument above to the case of
gravity coupled to scalars and vectors. In this way we shall 
vindicate the  claim made in \cite{GKK} that the mass
of a black hole with given charges is never less
than the double extreme hole with the same charges. 
For more details about attractors, double extreme holes etc and references to
the earlier literature the reader is directed to \cite{GKK,FGK}.

Consider a theory with matter lagrangian
\ben
L= -{ \over 2} (\partial \phi)^2 - { 1\over 16 \pi} e^{-2\phi} F^2.
\een
In general the scalar $\phi$ will take
a particular limiting value $\phi^{\infty}$ at infinity
and will vary over the interior. For static extreme 
holes it must take a particular value, called in the sequel
$\phi^{\rm frozen}$ on the horizon.
Static extreme black holes for which $\phi$ takes this value everywhere
are said to be double extreme and the scalar is said to be frozen.
They have the same geometry as the extreme Reissner-Nordstrom
black hole.
The mass of any regular static black hole with give charges $(Q,P)$
is never smalller than the value it takes $|Z(q,p)|=| \sqrt{ (2|QP|)}|$
for the double extreme hole with those charges. Since the area of the horizon of a double extreme hole is $A=2 \pi |PQ|$
One expects
that this should be a general lower bound for any time symmetric
 initial data. In fact we shall show that
 \ben
2M \ge {r_H^3}+ r_H + { Z^2 \over r_H^2} . 
\een
A simple calculation reveals that in the time
symmetic case
\ben
T_{00}= { 1\over 2} ( \nabla \phi)  ^2 +  { 1\over 8 \pi} e^{-2 \phi} 
( {\bf E} ^2 + {\bf B} ^2 ) .
\een
However now
\ben
{\nabla}. {\bf B}=0={\nabla}.{\bf D}.
\een
where the electric induction $\bf D$ is given by
\ben
{\bf D}= e^{-2 \phi} {\bf E}.
\een

Moreover now 
\ben
Q= { 1 \over 4\pi } \int_{S_s} D_n dA
\een
where $D_n$ is the normal component
of the electric induction  $\bf D$. The expression for the magnetic charge
remains unchanged.

An application of the Schwartz inequality now gives

\ben
 \int_{S_s} 16 \pi T_{00} \ge 2 { 1  \over A^2(s)} \int_{S_s} ( Q^2 e^{2\phi}
 + P^2 e^{-2\phi} )dA.
\een
Let's define:
\ben
V(\phi, Q,P)= ( Q^2 e^{2\phi}
 + P+2 e^{-2\phi} ).
\een
At fixed charges $(Q,P)$, both assumed to be non-vanishing,
 the function $V(\phi,Q,P)$ attains its least value 
$Z^2(Q,P)=2|PQ|$ at the so-called frozen value
$\phi= \phi ^{\rm frozen}$ given by
\ben
\phi^{\rm frozen}= { 1\over 2} \ln |P/Q|.
\een

Thus
\ben
 \int_{S_s} 16 \pi T_{00} \ge 32 \pi^2{ Z^2(Q,P)  \over A(s)}.
\een

The result now follows as in the Resissner-Nordstrom case.

We have chosen as simple example but it is clear that it extends
to include the general class of theories with any number of abelian
vector and scalar
fields coupled to gravity with an action which is quadratic in the vectors.

\section{ The second variation}

To conclude, we turn to the information about
 the area of a stable minimal 2-surface
in a time symmetric slice which comes from the second variation
of the area, i.e. the Hessian of the area functional,
 and demanding that it be positive.
The method is standard. Its application
in the
electrovac case goes back at least to \cite{GA}.  
For a one parameter variations $S_t=S+htn^i$ 
by an amount $h$ along the normals $n^i$ one has:

\ben
{ d^2 A \over dt ^2} = -\int _S h^2 \bigl [ ^3R_{ij} n^in^j -2 \sigma ^2 \bigr ] + \int _S |\nabla h|^2
\een
where the integral is taken over the minimal surface $S$
and $\sigma^2 ={ 1\over 2} \sigma_{ab} \sigma ^{ab}$ is the magnitude of the trace free
part of the second fundamental form $p_{ab}$ of the minimal surface $S$. 
The Gauss-Codazzi equation gives
\ben
2K_G=^3R-2^3R_{ij}n^i n^j -2 \sigma^2.
\een
Therefore:
\ben
{ d^2 A \over dt ^2} = \int  \bigl [|\nabla h|^2 + K_G h^2
- { 1\over 2} ^3Rh^2   -\sigma^2 h^2 \bigr ].
\een

If $h$ is taken to be constant over the surface 
and we use the Gauss-Bonnet theorem then we must have:
\ben
4\pi (1-g) -\Lambda A- \int_S  {8 \pi} T_{00} >0 ,
\label{variation}
\een
where the integral is taken over the minimal surface $S$.

If the cosmological constant is positive 
and the positive energy condition holds
we see a stable minimal surface must have spherical topology,
an old result. Moreover it's area must exceed $4 \pi \over \Lambda$.

If the cosmological constant is positive there will be
a cosmological horizon. This is not stable since its area decreases if
it is moved uniformly outward or inward. If that is the only negative mode
for the second variation one says that the surface is of index one.
Thus the second variation should be positive if 
\ben
\int_S h=0.
\een
A spectral argument explained to me by S.T. Yau
gives the the following lower bound for a suface of genus $g$
\ben
\int _S \bigl [ |\nabla h |^2 + K_G h^2 \bigr ] \Big / \int_S h^2  {\le  4 \pi (3-g) /A}.  
\een
Thus if $g=0$ we get:
\ben
A_C \le {12 \pi \over \Lambda}.
\een
This upper bound also follows from Geroch's method.
If the inverse mean curvature flow
starts from a black hole horizon of area $A_H$ and reaches a cosmological horizon of area $A_C$ ,
one obtains: 
\ben
\sqrt{ A_H }({12 \over \Lambda}- A_H) \le \sqrt{ A_C }({12 \over \Lambda}- A_C).
\een

( see \cite{MKNI} for a recent discussion of related results for 
positive cosmological constant).

If the  cosmological constant is negative and $k=-1$  
and the positive energy condition holds then 
the second variation yields
 a lower bound:
\ben
A >{4\pi (g-1) \over |\Lambda |}
\een

Thus there is a universal topology dependent {\sl lower} bound for the
entropy of a topological black hole.

If we are dealing with scalars and vectors, as in the previous
section we know that:
\ben
\int_S T_{00} \ge 2 \pi^2 { Z^2(Q,P) \over A}.
\een

This allows a strengthening of the bounds above.
Consider the case of black holes with spherical topology.
If the  cosmological constant is zero then 
we have a lower bound for the entropy with fixed charges,
valid for any time-symmetric initial data set:
\ben
A > 4 \pi Z^2.
\een

Again, this substantiates and extends to the non-spherically symmetric case the claims made in\cite{GKK}.

The results of this and previous sections strongly suport the idea that
\ben
S= \pi Z^2
\een
is the irreducible ammount of entropy of a black hole with fixed
charges $(Q,P)$. This irreducible amount of entropy
is only attained for extreme black holes. Any additional
gravitational  fields can only increase it.
This is consistent with the microscopic derivations
from D-brane theory but still leaves open the question of the 
origin of the extra contribution to the entropy
in the case of non-extreme states. Microscopically
this may be due to excited states of D-branes but
the  extrapolation  to macroscopic geometric
configrations of classical  initial data in the non-BPS
situations is not all all obvious.

\section{ The Horowitz-Myers conjecture}

One motivation for this present work was to see if one might use
the inverse mean curvature flow to tackle an interesting conjecture of 
Horowtiz and Myer's \cite{HM}. 
This was made in the context of the $AdS_5/CFT$ correspondence 
which concerns's 4+1 dimensional spacetimes. 
Thus the mean curvature flow is not immediately applicable.
However there is a version in $p+1$ dimensions
for all $p$, including $p=3$. Unfortunately
it turns out that the mean curvature flow is not 
applicable in that case either. 
To see why, we note that the Horowitz-Myer's example is 
obtained by interchanging the role of time with one of the torus
coordinates in the $k=0$ Kottler metrics:
\ben
ds^2= -{r ^2 \over a^2}dt^2 + {dr^2 \over (  {r^2 \over a^2}  -{r_0^3  \over a^2r} )}
+ (  {r^2 \over a^2} -{r_0^3  \over a^2r} ) dx ^2 + { r^2 \over a^2 }dy^2.
\een

Regularity demands that we identify $x$ with period $L_x={ 4 \pi a^2 \over 3 r_0}$. The period of
the coordinate $y$, call it $L_y$  is arbitrary. 
The initial data set thus has
topology $ {\Bbb R} ^2 \times S^1$. 
Note that the spacetime has no event horizon.
Horowitz and Myers' 
show that the ADM mass (obtained by comparing with the case $r_0=0$)
is negative and equals 
\ben
-{4 \pi ^2 L_y a^2 \over 27 L_x^2}.
\een

They give evidence to support the conjecture that this is a
lower bound for the ADM mass of such data sets, specified by $L_x$ and $L_y$.
As in the case of the negative mass $k=1-$ black holes
this suggestion is striking because the lower bound is attained
for a non-BPS state.

The difficulty with applying the inverse mean curvature flow is that
because the spacetime has no event horizon, the initial 
data set need not have, and indeed doesn't have, an  apparent horizon.
Thus there is no natural 2-surface from which to start the flow.
There is a minininal geodesic of length $r_0L_y$
at $r=r_0$ but this is no good because
if we consider a small  tube of radius $r$
surrounding the geodesic the function $f$ 
would tend to minus infinity like $r^{-1}$ as we shrunk the tube onto the geodesic. Since the area of the tube shrinks as $r$ the product $\sqrt{A} f$
diverges like $r^{-1\over 2}$. The monotonicity poperty thus seems to 
give nothing interesting.
On the other had if we were to start the flow from a very small sphere
we would, assuming that the flow reaches infinity, merely obtain the result
that the quantity $r_0$ is positive.

\section {Cosmic Censorship vesus Bogomol'nyi}

If $\Lambda <0$ and $k=1$ we obtain the Cosmic Censorship lower bound:
\ben
M\ge r_H + {r_H^2 \over a^2}  + {Z^2 \over r_H}.
\een
By extremizing with respect to $r_H$ we find, after some 
elementary algebra and manipulation of surds,
the following  interesting the lower bound for $M$ in terms of $ |Z|$ and
\ben
M \ge  \sqrt {2 \over 3}|Z|  
\Bigl [
\sqrt{1 + \sqrt {1-4\Lambda Z^2} }
+{ 1 \over 
\sqrt{1 + \sqrt {1-4\Lambda Z^2} }}
\Bigr ] 
\label{exotic}
\een

Of course in the limit that $\Lambda \rightarrow 0$ we recover the
familiar Bogomol'nyi bound:

\ben
M \ge |Z|, \label{standard}
\een

but it is clear  that the Cosmic Censorship lower bound (\ref{exotic}) is 
strictly greater than the Bogomolnyi lower bound (\ref{standard})
if $\Lambda <0$.

\section{Higher dimensions}

As originally formulated the Geroch technique 
works only for three spatial dimensional initial data sets.
However there is an analogue of the positive
mass theorem in all higher dimensions. This has been proved
by Schoen and Yau and by Wiitten. 
A reformulation of Geroch's idea designed to see
how this might work
in higher dimensions
(shown to me by Douglas Eardley in 1980) goes as follows:
One considers a foliation of an $n$-dimensional
manifold of the form:
\ben
ds^2=e^{2\psi} dr^2+r^2 h_{ab}(x,r)dx^a dx^b
\een
where 
\ben
{\partial \over \partial r} {\rm det} h_{ab} =0,\een
Thus the trace of the second fundamental form is
\ben
p={n-1 \over r}.
\een
Let 
\ben
q_{ab}= { 1\over 2} e^{-\psi} {\partial \over \partial r} h_{ab}\een
so that
\ben
q_{ab} = { 1\over r^2}( p_{ab}-{ 1\over n-1} h_{ab} p) 
\een
The Ricci scalar is given by
\ben
^n R= { 1\over r^{n-1}} {\partial \over \partial r}
 \bigl [ (n-1) r^{n-2} ( 1-e^{ -2\psi} ) \bigr ] - { 2\over r^2}
e^{-\psi} D^a D_a e^{\psi} - q^{ab} q_{ab} +
 { 1\over r^2} \bigl [ ^{n-1}R-(n-1)(n-2) \bigl ].
\een
where $ ^{n-1}R $ is the scalar curvature of the 
the level sets $r={\rm constant}$.

One may now multiply this equation by
$r^{n-1} \sqrt{{\rm det} h_{ab}} dr d^{n-1}x $ and integrate
to get the relevant identities. 
One assumes that the outermost aparrent horizon
is located at $r=r_H$ on which $\psi=0$.
If the metric is asymptotically flat,
it is convenient to normalize such that
\ben
\int _{S_r} \sqrt{{\rm det} h_{ab}} d^{n-1}x= {\rm vol}(S^{n-1}),
\een
where ${\rm vol}(S^{n-1}) $ is the volume of a standard round sphere
of unit radius. Then at large distances
\ben
\psi \sim { a \over 2 r^{n-2}}
\een
where the total mass $M$ is given by
\ben
M={ (n-1) {\rm vol}(S^{n-1}) \over 16 \pi} a
\een

We get the integral inequality
\ben
a\ge r^{n-2} _H +\int F r^{n-3} dr
\een
where
\ben
F(r)=\int _{S_r} \sqrt{{\rm det} h_{ab}} d^{n-1}x  \bigl [ ^{n-1}R-(n-2)(n-1) \bigl ].
\een
To make the theorems work we need this to be non-positive.
If $n=3$ this follows from the Gauss-Bonnet theorem. For $n>3$
the situation is less clear. The integrand vanishes for
a round sphere and this fact leads to some obvious
positivity properties if we assume that the
levels sets are the orbits of an 
$SO(n)$ isometry group. If the metric is not $SO(n)$-invariant however
we have, without further information, little control of the integrand.
The quantity $F$  is of course related  to to the Yamabe constant.
The fact that this has already arisen in the context of entropy
in cosmology \cite{GG3,GG4} is striking. 

Note that to derive inequalities from
the second variation(\ref{variation}) we needed 
to use the Gauss-Bonnet theorem.
In higher dimensions what appears in the second variation
is the quantity

\ben 
\int_S R
\een
where $R$ is the Ricci scalar of the apparent horizon.
If the comological constant is non-negative this must be postive.
In fact the Ricci scalar
 should probably be point wise positive. This would place some
topological restrictions on the topology of
the apparent horizon. 

Thus it seems that the only real
evidence for a bound to the area of apparent
horizons comes from the higher dimensional versions of
the collapsing shell calculations \cite{GG5}.

\section{Acknowledgement}
I should like to thank Gary Horowitz and Tom Ilmanen for a number of helpful
discussions.

\end{document}